\documentclass[conference]{IEEEtran}
\usepackage{todonotes}            
\presetkeys{todonotes}{inline}{}  
\usepackage{tabularx,multirow} 
\usepackage[binary-units=true]{siunitx}
\usepackage[capitalise]{cleveref} 

\usepackage{cite}

 \usepackage{eso-pic}
 \usepackage{color}
 \definecolor{DisclaimerGray}{gray}{0.92}
 
 \AddToShipoutPicture*{\put(45,760){\colorbox{DisclaimerGray}{\parbox{\textwidth}
 {\footnotesize \copyright 2020 IEEE.  Personal use of this material is permitted.  Permission from IEEE must be obtained for all other uses, in any current or future media, including reprinting/republishing this material for advertising or promotional purposes, creating new collective works, for resale or redistribution to servers or lists, or reuse of any copyrighted component of this work in other works. The definite version will be published at  
2020 IEEE Power \& Energy Society Innovative Smart Grid Technologies Conference (ISGT).}}}}

\ifCLASSINFOpdf
\else
\fi
\usepackage{amsmath}
\usepackage{url}

\title{Towards Comparability in Non-Intrusive Load Monitoring: On Data and Performance Evaluation}

\author{\IEEEauthorblockN{Christoph Klemenjak}
\IEEEauthorblockA{University of Klagenfurt\\
Klagenfurt, Austria\\
\url{klemenjak@ieee.org}}
\and
\IEEEauthorblockN{Stephen Makonin}
\IEEEauthorblockA{Simon Fraser University\\
Burnaby, Canada\\
smakonin@sfu.ca}
\and
\IEEEauthorblockN{Wilfried Elmenreich}
\IEEEauthorblockA{University of Klagenfurt\\
Klagenfurt, Austria\\
wilfried.elmenreich@aau.at}}

\begin{document}

\maketitle

\begin{abstract}
Non-Intrusive Load Monitoring (NILM) comprises of a set of techniques that provide insights into the energy consumption of households and industrial facilities. Latest contributions show significant improvements in terms of accuracy and generalisation abilities. Despite all progress made concerning disaggregation techniques, performance evaluation and comparability remains an open research question. The lack of standardisation and consensus on evaluation procedures makes reproducibility and comparability extremely difficult.
In this paper, we draw attention to comparability in NILM with a focus on highlighting the considerable differences amongst common energy datasets used to test the performance  of algorithms. We divide discussion on comparability into data aspects, performance metrics, and give a close view on evaluation processes. Detailed information on pre-processing as well as data cleaning methods, the importance of unified performance reporting, and the need for complexity measures in load disaggregation are found to be the most urgent issues in NILM-related research.  In addition, our evaluation suggests that datasets should be chosen carefully. We conclude by formulating suggestions for future work to enhance comparability.
\end{abstract}


%
\IEEEpeerreviewmaketitle

\section{Introduction} \label{sec:intro}

Non-intrusive load monitoring, also referred to as load disaggregation, originates from Hart's~\cite{hart1992nonintrusive} seminal work and comprises a set of techniques that provide deep insights into energy consumption of buildings, detached houses, and apartments based on smart meter data.
NILM techniques enable, inter alia, occupancy detection for health-monitoring purposes, optimisation of workflows inside industrial facilities, and help in achieving cost reduction by providing immediate detailed feedback about a user's energy consumption \cite{klemenjak2018yomopie}.
Latest NILM contributions present significant improvements as a result of successful integration of Deep Learning \cite{Nalmpantis2018}, \cite{murray2019transferability}.

However, with an arising discussion of transferability -- using one dataset to train and a different dataset to test -- in NILM \cite{murray2019transferability}, a widely unsolved research issue gains importance: \textit{comparability in NILM research}.
The comprehensive review on performance evaluation in \cite{pereira2018performance} points out that there is no consensus regarding what metrics should be used and identify the absence of a formal agreement on how to report evaluation results in load disaggregation scholarship. In the same vein, studies of machine learning approaches for NILM find that as a consequence of the variety of datasets and the diversity of methodologies, an objective comparison is almost impossible at the moment \cite{herrero2017non}, \cite{Nalmpantis2018}.
Furthermore, it is not fully understood how disaggregation problems can be compared and what properties of a dataset influence the problem's complexity \cite{egarter2015complexity}.


In this paper, we investigate differences of common \textit{low sampling rate} energy datasets and initiate discussion on a widely disregarded issue in NILM, \emph{comparability}. 
We divide our discussion on comparability into aspects with regard to datasets, metrics, and the evaluation process itself. Finally, we conclude the paper with formulating recommendations for future work.

It should be noted that \textit{low sampling rate} is defined as data sampled at rates between 1 to 60 samples per minute, inclusively.
All source code used in the course of our investigations is compatible with NILMTK \cite{batra2014nilmtk}. More information as well as supplemental material can be obtained from our repository\footnote{\url{https://github.com/klemenjak/comparability}}.


\section{A Definition of NILM} \label{sec:definition}

In the previous section we define NILM conceptually. Mathematically, we describe NILM as the problem of providing estimates $[\hat{x}_t^{(1)}, \dots ,\hat{x}_t^{(M)}]$ of the actual power consumption of $M$ electrical appliances $[x_t^{(1)}, \dots ,x_t^{(M)}]$ at time $t$ given only the aggregated power consumption $y_t$.
We refer to algorithms applied to solve load disaggregation problems as load disaggregation algorithms.
The aggregate power signal $y_t$, provided to a load disaggregation algorithm consists of
\begin{equation}
	y_t = \epsilon_t + \sum_{i=1}^{M}{x_t^{(i)}}
\end{equation}
$M$ appliance-level signals $x_t^{(i)}$ and an error term $\epsilon_t$, which models the discrepancy between the sum of the individual measurements and the overall branch measurement. The error term consists of (measurement) noise and unmetered appliance-level signals.

\renewcommand{\arraystretch}{1.3}
\begin{table*}[t]
\begin{center}

  \caption{A comparison of selected households embedded in common energy datasets}
  \label{tab:comparison}
\begin{tabular}{cccccccccccccccccccc}
\hline
Dataset &  & House &  & Duration   &  & Meters &  & \multicolumn{2}{c}{Sampling} &  & \multicolumn{2}{c}{AC Power Types} &  & \multicolumn{2}{c}{Number of Events} &  & \multicolumn{2}{c}{NAR} \\
        &  &       &  &            &  &        &  & Mains         & Sub          &  & Mains           & Sub           &  & Min           & Avg          &  & P          & S          \\
        &  &       &  & {[}days{]} &  &        &  & {[}s{]}       & {[}s{]}      &  &                 &               &  &               &              &  & {[}\%{]}   & {[}\%{]}   \\ \hline
        
AMPds2     &  &   1 of 1    &  & 730                    &  &  20    &  &     60    & 60     &  &   P, Q, S    &    P, Q, S   &  &   0 & 319   &    &   18    &     6     \\

    COMBED       &  &   1 of 2   &  & 28                    &  &  13    &  &     30    & 30     &  &   P   &    P  &  &   0 & 463 &  &       34    &     -       \\
    
DRED    &  &   1 of 1   &  & 153                    &  &  12    &  &     1    & 1     &  &   S              &    S &  &  1  & 604  &  &       -    &     28       \\

ECO    &  &   1 of 6   &  & 245                    &  &  7    &  &     1    & 1     &  &   P, Q             &    P &  &   7 & 691  &  &   68    &     -      \\

ECO   &  &   6 of 6   &  & 219                    &  &  7    &  &     1    & 1     &  &   P, Q             &    P &  &  1 & 1166   &  &   74    &     -      \\
 
iAWE    &  &   1 of 1   &  & 73                    &  &  10    &  &     1    & 6     &  &   P, Q, S             &    P, Q, S &  &   1 & 497  &  &   63    &     61      \\

REDD    &  &   1 of 6   &  & 36                    &  &  16    &  &     1    & 3     &  &   S            &    P &  &   0 & 799 &  &  -   &     -      \\

REDD    &  &   2 of 6   &  & 35                    &  &  9    &  &     1    & 3     &  &   S            &    P &  &   0 & 168 &  &  -   &     -      \\


REFIT    &  &   1 of 20   &  & 638                    &  &  9   &  &     7    & 7     &  &   P            &    P&  &   1 & 320  &  &  65   &     -      \\

REFIT    &  &   8 of 20   &  & 555                    &  &  9    &  &     7    & 7    &  &   P           &    P &  &   3 & 229 &  &  78   &     -      \\

REFIT    &  &   17 of 20   &  & 443                    &  &  9    &  &     7    & 7   &  &   P            &    P&  &   1 & 379  &  &  45   &     -      \\

UK-DALE     &  &   1 of 5   &  & 658                    &  &  52    &  &     1    & 6   &  &   P,S            &    P, S &  &   0 & 874 &  &  33  &  87  \\

UK-DALE   &  &   2 of 5   &  & 176                    &  &  18    &  &     1    & 6   &  &   P, S     &    P &  &   0 & 733 &  &  41  &  - \\

UK-DALE   &  &   5 of 5   &  & 137                    &  &  24    &  &     1    & 6   &  &   P, S           &  P&  &   0 & 1320  &  &  31  &  -  \\ \hline

\end{tabular}
\end{center}
\end{table*}

\section{Brief Comparison of Datasets} \label{sec:datasets}

Real-world energy datasets are crucial for the development and testing of signal processing and machine learning algorithms to solve energy related problems such as load disaggregation \cite{klemenjak2019datasets}. Such datasets are the outcome of measurement campaigns in households and/or industrial facilities with special attention to not disrupt everyday routines within the monitored space so that the recorded dataset resembles reality as best as possible.
It is common practice to test novel approaches on several datasets to demonstrate versatility as well as generalisation abilities \cite{pereira2018performance}.

Our dataset analysis considered several public low-sampling-rate energy datasets usable in NILMTK. 
Table~\ref{tab:comparison} summarises the outcome of our analysis. We compare selected households embedded in AMPds2~\cite{makonin2016ampds}, COMBED~\cite{batra2014comparison}, DRED~\cite{uttama2015loced}, ECO~\cite{beckel2014eco}, iAWE~\cite{batra2013s}, REDD~\cite{kolter2011redd}, REFIT~\cite{refit15}, and UK-DALE~\cite{kelly2015ukdale}.
BLUED was excluded from our investigations due to the lack of sub-metered power data, Tracebase and GREEND due to the lack of household aggregate power data \cite{batra2014nilmtk}.


\subsection{Measurement Campaigns}

Since 2011, there has been an increase in datasets recorded around the world. A comprehensive overview can be obtained from \cite{pereira2018performance}. Conducted measurement campaigns share common aims per se, namely recording energy consumption and other parameters of interest in selected households over a certain period of time. However, we can observe considerable differences in the way past campaigns have been conducted.
As Table \ref{tab:comparison} shows, campaign durations differ significantly ranging from a couple of days to several years of data, which impacts the number of appliance activations and events caught. Another noticeable difference between existing datasets lies in campaign scaling. We can identify many small-scale campaigns covering a small number of households but also some large-scale campaigns that incorporate up to 20 households.
With regard to measurement setups, datasets depicted in Table \ref{tab:comparison} show large variations in terms of available AC power types, sampling rates, and the number of installed sub-meters. Furthermore, it should be noted that there seems to be a lack of consistency in the sense that not only measurement setups between two datasets differ significantly but also setups within some considered campaigns. 


\subsection{Number of Events}

Comparisons between monitored households reveal significant differences of individual habits and daily routines of occupants. These habits and individual routines affect the usage of household appliances and therefore, the number of events found in energy consumption data. We hypothesise that the number of events has a considerable impact on the performance of load disaggregation algorithms since a high number of observed events would reflect a vibrant household. In Table \ref{tab:comparison}, we summarise the number of events detected in selected households of commonly-used energy datasets in NILM scholarship.
For each individual appliance of a respective household, we estimate the number of events embedded in the appliance's power consumption trace. In this context, we define an event to be the transition between two representative states of power consumption. These representative states are obtained by applying methods of statistics, filtering, and clustering. A detailed description can be obtained from the supplemental material.

Table \ref{tab:comparison} reports statistics related to the number of events per day. The \emph{minimum number of events} provides information on the least-active appliance of a household i.e. the appliance with the lowest average of events per day. For a considerable number of households considered in our study, this minimum equals 0. From this follows that these households contain records of one or more appliances, which have not been used at all during the measurement campaign and therefore, the assigned meters have recorded only noise. 
In addition, we report the \emph{average number of events per day}. This average provides information on how many events can be observed in a respective household on average per day. In this way, it's possible to compare the level of activity in households and possibly also to draw conclusions on the difficulty of detecting events. We observe a significant difference for this measure, not only between different datasets but also between households of the same dataset.







\subsection{Noise Level}

For practical reasons, not every single electrical appliance can be equipped with a measurement instrument during a measurement campaign. For instance, attaching a meter to a water heater represents a challenge, when accessing the switchboard is not possible or prohibited. 
Consequently, the aggregate power signal $y_t$ of a real-world dataset consists not exclusively of known appliance-level signals $x_t^{(i)}$, but also contains several unknown appliance-level signals that contribute to the error term $\epsilon_t$.
Therefore, there exists a direct link between the number of installed sub-meters and the amount of unknown and unwanted components contributing to the aggregate power signal.
To quantify the amount of noise of an aggregate power signal, the authors of \cite{makonin2015nonintrusive} introduced the percent-noisy measure (\%-NM). We adopt this measure and refer to it as \emph{noise-to-aggregate ratio (NAR)}:
\begin{equation}
	\text{NAR} = \frac{\sum_{t=1}^{T}{|y_t-\sum_{i=1}^{M}{x_t^{(i)}}| }}{\sum_{t=1}^{T}{y_t}} \end{equation}
where $x_t^{(i)}$ is the power consumption of appliance $i$, $y_t$ the aggregate power signal, and $T$ the length of the observed time frame. 
The noise-to-aggregate ratio (NAR) can be computed for all AC power types, as long as energy readings of aggregate and sub-meters are available.
A ratio of 0.25 reports that 25\% of the total energy consumption stems from unmetered appliances and noise. Hence, the ratio indicates to what degree information on the aggregate's components is available.
In Table~\ref{tab:comparison}, we summarise the noise-to-aggregate ratio (NAR) for selected households with regard to active power P and apparent power S.
In general, a high number of installed sub-meters results in a low NAR, as can be observed in dataset AMPds2. It should be noted that appliance types play an important role in this matter. Appliances such as electric stoves, water heaters, and clothes washers consume considerably more energy than electronic low-power gadgets. Hence, considering them during a measurement campaign eventually reduces the final NAR.
In our comparison, several households show a NAR higher than 50\%. From this follows that we have very limited knowledge about the aggregate's composition. Likely causes for such a high ratio are a low number of installed measurement instruments or a poor selection of monitored appliances.


\section{Comparability in NILM} \label{sec:comparability}

Comparability in the context of load disaggregation is a multifaceted issue that comprises dataset aspects, performance metrics, disaggregation techniques, and aspects related to the evaluation process. We review comparability in NILM by breaking down this complex matter into individual aspects: data, accuracy metrics, and performance evaluation.


\subsection{Data} 

Marginal research efforts have been spent in order to clearly understand important \emph{properties of energy datasets} and their impact on the performance of load disaggregation algorithms. As the results depicted in Table \ref{tab:comparison} indicate, there are significant differences between energy datasets. 
Therefore, we claim that treating datasets to be interchangeable for evaluating algorithms can be misleading. Furthermore the individual characteristics of a dataset have a decisive influence on the performance of load disaggregation techniques. Yet, it is not understood what are the relevant characteristic properties of a dataset and how to quantify them. Therefore, we identify an urgent need for measures that enable meaningful comparisons of energy datasets. This is of special interest in cases where closed datasets are used, which cannot be published as a result of non-disclosure agreements or privacy concerns. With regard to privacy, appropriate measures could serve to protect privacy and enable comparability of datasets at the same time.
The theoretical considerations on a complexity measure for NILM in \cite{egarter2015complexity} represent a first approach towards quantifying the complexity of datasets. As the authors state, such measures need to be independent of load disaggregation approaches while still taking into account a variety of factors such as the number of appliances, appliance types, and similarity of appliance states. We claim that profound research on these topics has to be conducted in order to overcome comparability issues.

\subsection{Accuracy Metrics}


In load disaggregation, performance evaluation aims to assess the effectiveness of a method by comparing the observed appliance signal (ground-truth) and provided estimates. Despite a big variety of performance measures can be observed in related work, it is crucial to select metrics carefully in order to avoid misinterpretations of results \cite{Mayhorn2016}. Prevalent metrics utilised in NILM can be divided into event detection and energy estimation metrics \cite{Mayhorn2016}, \cite{pereira2018performance}. 
The authors of \cite{Mayhorn2016} examined selected metrics for their effectiveness in quantifying disaggregation performance. The study finds that the metrics energy error, energy accuracy, and match rate are best suited. 
In \cite{Pereira2017}, researchers compare 18 performance metrics for event classification. The behaviour is compared when applied to classification algorithms in event-based load disaggregation. Conducted studies show high correlations between most event-based metrics. The authors also find that probabilistic measures can provide information that is not available when using more traditional metrics. 

With regard to comparability, we claim that there is no common understanding or accepted format (i.e., unified reporting approach) as to how to report on testing setup and accuracy results such as the one presented in \cite{makonin2015nonintrusive}. As a consequence of the variety of existing load disaggregation techniques, performance evaluation has to assess classification performance as well as performance related to energy estimation in order to enable comparability.
Furthermore, we recommend the usage of normalised metrics, so that low-power appliances and appliances with a significant power consumption can be compared in a substantial manner.

As we observe an upcoming discussion on transferability of NILM approaches, as in \cite{murray2019transferability}, the question arises if performance evaluation should look beyond accuracy and consider additional properties of a NILM algorithm such as generalisation abilities, scalability, and privacy-preserving features. We claim that the ability of a NILM algorithm to deliver good performance on unseen scenarios is certainly in the interest of consumers and therefore, should be assessed.

\subsection{Performance Evaluation}

\begin{table}[]
  \caption{RMSE for different test set ratios (TSR) on REDD}
  \label{tab:tsr_redd}
  \begin{center}
      
\begin{tabular}{llrllll}
\hline
\multicolumn{1}{c}{} & \multicolumn{2}{c}{TSR = 25.9\%} & \multicolumn{2}{c}{TSR = 17.1\%} & \multicolumn{2}{c}{TSR = 8.5\%} \\
appliance            & CO            & FHMM           & CO           & FHMM           & CO         & FHMM         \\\hline
light  &    224.3    &    47.5 & 210.1 & 43.9 &      183.2    &  43.7             \\
fridge    &     98.9  &   99.2 & 94.4 & 82.4    & 79.1    &   73.0               \\
sockets &   116.9    &  148.6  & 104.8 & 136.7        &   100.1  & 125.4             \\
el. heater &   126.0        &  141.0   & 96.7 & 126.0  & 72.4     & 96.5              \\
AC &   231.3    &  227.6   & 74.2 & 65.3       & 41.2     &   44.1             \\ \hline       
\end{tabular}
  \end{center}
\end{table}


It seems to be common practice that researchers evaluate NILM solutions on different datasets, with different criteria, and with the help of different metrics \cite{Nalmpantis2018}. Domain-specific tool kits for performance evaluation in NILM exist \cite{batra2014nilmtk}, \cite{beckel2014eco}, but the absence of standardisation of evaluation procedures results in comparability remaining an open issue. Pre-processing methods, the extent of testing, and whether testing is performed on denoised or real test sets greatly influences the outcome of evaluations in NILM scholarship.

As concerns pre-processing of datasets, we claim that scholars should describe precisely any manipulations of datasets carried out before evaluation. Re-sampling, data cleaning methods, dataset balancing, and bias countermeasures significantly alter the characteristics of datasets. Consequently, comparing the outcome to its initial shape would be misleading and relativise the advantage of having common public datasets \cite{makonin2015nonintrusive}. 

In order to obtain conclusive results in performance evaluation, NILM algorithms have to be tested on a sufficiently large amount of data. However, we can observe large variations in related work spanning from a few days up to several months of test sets.
We identify the need for a simple measure that gives information on how extensively testing was performed on a dataset.
We suggest reporting the amount of data used for evaluation to get an idea of how many events, i.e., appliance transitions~\cite{Pereira2017}, were embedded in the test set.
To quantify this property, we propose the \emph{test set ratio (TSR)} and the \emph{event ratio (EVR)}, which are defined as

\begin{equation}
    \text{TSR} = \frac{\text{test duration}}{\text{total duration}}  \quad\text{and}\quad \text{EVR} = \frac{\text{events in test set}}{\text{events in dataset}}
\end{equation}

the ratio between test duration and the total duration of a time series for energy estimation purposes and the ratio between the number of events in the test set and the total number of events in the dataset. In case of a significant amount of missing data intervals, e.g., a measurement that goes over a year and is missing a month, the duration will be calculated as the aggregation of of all sub-durations.

With these metrics, we are able to put into relation evaluation results and test set size. We argue that this is an appropriate approach towards comparability since not all chunks of a dataset have equivalent properties. For example, a dataset might include some events that are difficult to detect or to distinguish while another part may include items that have clear characteristics. Therefore, testing on a high number of chunks should be favoured over "cherry-picking" of single chunks. 
To point out the importance of metrics such as TSR or EVR, we present a case study on the dataset REDD. We trained and evaluated two of NILMTK's disaggregation algorithms, one based on Combinatorial Optimisation (CO) and the other one using Factorial Hidden Markov Models (FHMM). We extracted three testsets from house 6 with considerably different test set ratios (TSR) of 25.9\%, 17.1\%, and 8.5\%. In Table \ref{tab:tsr_redd}, we summarises the disaggregation error for our three disaggregation studies. We used the same training set in all three studies. To quantify the disaggregation error, we utilise the root-mean-squared error (RMSE) between the ground-truth signal $x_i$ and the estimated power consumption $x_i$ of appliance $i$.

\begin{equation}
    \text{RMSE} = \sqrt{\frac{1}{T} \cdot \sum_{t=0}^{T-1}{(\hat{x}_t^{(i)}-x_t^{(i)})^2}}
\end{equation}

We observe a significant lower disaggregation error for the testset with a TSR of 8.5\%, when we compare the results of our two disaggregation algorithms to the testset with a TSR of 25.9\%. In general, we identify smaller disaggregation errors for smaller testsets. Also, we see that the TSR metric can assist in pointing out that evaluation has been performed on a very small subset and therefore, incentivise selecting larger testsets.



Evaluations in NILM can be carried out either in a noised or denoised manner. The difference lies in the composition of the aggregate power signal.
Whereas in noised testing, the aggregate power signal, for instance, stems from a smart meter, the aggregate signal in denoised settings is an artificial composition of known appliance level signals. In general, such denoised settings result in higher accuracies than real-world settings \cite{makonin2015nonintrusive}. Experiments presented in \cite{bonfigli2018denoising} support this assertion, where the performance of the AFAMAP algorithm and denoising autoencoders was evaluated on noised and denoised scenarios with noticeable differences in performance. 

However, denoised evaluation settings do not reflect real-world situations since assuming to have knowledge about all components of an aggregate signal is rather idealistic.
For this reason, we recommend reporting the amount of noise of a test set using the noise-to-aggregate ratio (NAR). This represents a simple and effective measure to distinguish between noised and denoised testing scenarios.
The following case study underlines why reporting the NAR is important. We trained and evaluated NILMTK's CO and FHMM algorithm on a subset of the DRED~\cite{uttama2015loced} dataset. We extracted the training set and a testset with a TSR of 35.2\% from house 1 of the dataset. We then trained both algorithms on DRED and disaggregated first the aggregate signal of the testset. For comparison, we disaggreated the denoised aggregate signal i.e. the sum of appliance signals found in the testset. As Table \ref{tab:dred} reports, we observe significant differences between the real-world and the denoised scenario with a maximum difference in disaggregation error of $\SI{14.2}{\watt}$ for the CO algorithm and $\SI{9.3}{\watt}$ for FHMM. As the results indicate, denoised testing yields a lower disaggregation error and, therefore, better evaluation results.

\begin{table}[]
  \caption{RMSE for normal and denoised testing on DRED} 
  \label{tab:dred}
\begin{center}
\begin{tabular}{llrllll}
\hline
\multicolumn{1}{c}{} & \multicolumn{2}{c}{NAR = 31.4\%} & \multicolumn{2}{c}{NAR = 0.0\%} & \multicolumn{2}{c}{difference} \\
appliance            & CO           & FHMM           & CO           & FHMM           & CO            & FHMM           \\\hline
cooker  &     46.3         &   45.5           &   39.2   &   38.4  & 7.1  & 7.1          \\
microwave  &    68.2       &      60.7          &    54.0       &    51.4 & 14.2 &  9.3          \\
laptop  &      19.0        &     18.0          &   16.9            &  13.8  & 2.1 & 4.2          \\
TV  &       29.6     &      27.0        &     24.2          &    23.5  & 5.4 & 3.5         \\
fridge  &      45.8       &  41.1            &   43.0            &   36.7  &2.8 & 4.4          \\
el.\ heater  &     32.1      &     28.2          &   27.0     &     27.5  & 5.1 & 0.7        \\ \hline            
\end{tabular}
\end{center}
\end{table}

\section{Conclusions} \label{sec:conclusions}

In this paper, we addressed certain domain-specific aspects relating to NILM. We compared common energy datasets and reported noteworthy differences between them in many regards. We identify an absent consensus on strategies for measurement campaigns and in particular, measurement setups. 

A further aim of this paper is to foster discussion on comparability in NILM with respect to dataset aspects, accuracy metrics, and performance evaluation. Recommendations on how to enhance comparability are provided which will be a basis for a further in-depth investigation. In particular, we suggest assessing the noise level in aggregate power signals and introduce two metrics for performance evaluation in NILM: EVR and TSR. Furthermore, we identified a strong need for unified quantitative complexity measures for load disaggregation datasets as a key issue to overcome comparability issues in future work.

\end{document}